\documentclass[%
    aip,                            
    amsmath,amssymb,
    reprint,    
    floatfix,
]{revtex4-1}

\usepackage[english]{babel}
\usepackage[T1]{fontenc}
\usepackage{textcomp} 				
\usepackage[dvips]{graphicx}		
\usepackage{color}					
\usepackage{verbatim}               
\usepackage{appendix}
\usepackage{ulem}
\usepackage{float}
\usepackage{siunitx}                

\renewcommand{\paragraph}[1]{{\bf #1}}

\graphicspath{ {figures/} }

\begin{document}

\title{Role of the ferroelastic strain in the optical absorption of BiVO$_4$}
\author{Christina Hill}
\affiliation{Materials Research and Technology Department, Luxembourg Institute of Science and Technology, 41 rue du Brill, L-4422 Belvaux, Luxembourg}
\affiliation{Department of Physics and Materials Science, University of Luxembourg, 41 rue du Brill, L-4422 Belvaux, Luxembourg}
\author{Mads C. Weber}
\author{Jannis Lehmann}
\author{Tariq Leinen}
\author{Manfred Fiebig}
\affiliation{Department of Materials, ETH Zurich, Vladimir-Prelog-Weg 4, 8093 Zurich, Switzerland}
\author{Jens Kreisel}
\author{Mael Guennou}
\affiliation{Department of Physics and Materials Science, University of Luxembourg, 41 rue du Brill, L-4422 Belvaux, Luxembourg}

\begin{abstract}
Bismuth vanadate (BiVO$_4$) has recently been under focus for its potential use in photocatalysis thanks to its well-suited absorption edge in the visible light range. Here, we characterize the optical absorption of a BiVO$_4$ single crystal as a function of temperature and polarization direction by reflectance and transmittance spectroscopy. The optical band gap is found to be very sensitive to temperature, and to the monoclinic-to-tetragonal ferroelastic transition at \SI{523}{\K}. The anisotropy, as measured by the difference in absorption edge for light polarized parallel and perpendicular to the principal axis, is reduced from \SI{0.2}{\eV} in the high-temperature tetragonal phase to \SI{0.1}{\eV} at ambient temperature. We show that this evolution is dominantly controlled by the ferroelastic shear strain. These findings provide a route for further optimization of bismuth-vanadate-based light absorbers in photocatalytic devices.


\end{abstract}

\maketitle


Bismuth vanadate (BiVO$_4$) has recently attracted considerable attention for photocatalytic applications.\cite{abdi_efficient_2013,bornoz_bismuth_2014,chen_reactive_2013,han_efficient_2014} This is primarily due to its optical absorption properties in the visible range: the reported band gap value\cite{cooper_indirect_2015} of \SI{2.5}{\eV} is close to the optimal value\cite{walsh_band_2009} of \SI{2.2}{\eV} desired for an efficient use of the solar spectrum. The position of the valence band edge of BiVO$_4$ is also well suited for efficient band-alignment for photoanodes in water splitting devices\cite{park_progress_2013}. Besides, bismuth vanadate exhibits high chemical- and photo-stability in aqueous electrolytes\cite{chu_roadmap_2017} and hence does not require a protective layer. To fully understand and optimize the behavior of BiVO$_4$ as a photoabsorber material, a detailed understanding of its electronic and optical properties is highly desirable. This task, however, is relatively complex compared to classical semiconducting materials, due to the chemical complexity of BiVO$_4$, its optical anisotropy, and its polymorphism.

BiVO$_4$ has three common crystal forms whose accessibility depends on the synthesis method:\cite{tokunaga_selective_2001} the zircon structure with tetragonal symmetry and two scheelite structures with monoclinic and tetragonal structure.\cite{tokunaga_selective_2001, das_investigation_2017} It is known that only the monoclinic scheelite BiVO$_4$ possesses optical properties suitable for applications in photocatalysis.\cite{zhang_selective_2007} In its scheelite structure, BiVO$_4$ exhibits a reversible second-order ferroelastic phase transition at $T$\textsubscript{c}$=523 \pm 5$\,K from a high-temperature tetragonal symmetry (space group I4$_1$/a) to a low-temperature monoclinic symmetry (space group I2/a -- C2/c(15) in the standard setting).\cite{bierlein_ferroelasticity_1975,david_structure_1979, sleight_crystal_1979, pellicer-porres_phase_2018} This phase transition is driven by a spontaneous shear strain developing in the plane perpendicular to the tetragonal axis; it was studied in detail in the past as a rather uncommon example of a proper ferroelastic transition.\cite{Benyuan1981,hazen_bismuth_1982, wood_ferroelastic_1980, david_ferroelastic_1983_III, david_ferroelastic_1983_IV, david_ferroelastic_1983_V,cummins1983} 

The electronic and optical properties of scheelite BiVO$_4$, in contrast, have been under focus much more recently, and are still under discussion. While the value of the band gap, around 2.5\,eV, is agreed on, its nature is controversial. On the one hand, several experimental works report a direct band gap.\cite{stoughton_adsorption-controlled_2013, luo_structural_2008, zhou_fabrication_2011} On the other hand, most density functional theory (DFT) studies conclude that the band gap is indirect, but also predict a direct transition only a few hundreds of meV above the indirect band gap.\cite{zhao_electronic_2011,das_investigation_2017} This scenario is difficult to confirm by classical methods, such as optical spectroscopy or ellipsometry. Only recently, Cooper et al.\ confirmed by resonant inelastic X-ray scattering that the fundamental band gap is indeed indirect with a direct band-to-band transition only \SI{200}{\meV} above it.\cite{cooper_indirect_2015} In addition, early single crystal studies have reported strong changes in color with varying temperature.\cite{david_ferroelastic_1983_V} This raises important questions on whether the direct and indirect band gaps show the same temperature dependence, and how reliable the comparison of experimental room temperature data with \SI{0}{\K}-DFT calculations is. Finally, because of the anisotropic structure of scheelite BiVO$_4$, dichroism is expected and has indeed been predicted by several groups through DFT calculations\cite{zhao_electronic_2011, ding_density_2013, das_investigation_2017} but has not be confirmed experimentally.

Here, we address these questions by a linear spectroscopy study of scheelite BiVO$_4$ single crystals as a function of temperature. We quantify the thermochromic behavior and the anisotropy of the optical absorption, and show that it is sensitive to the ferroelastic phase transition. Specifically, we show that, in the monolinic phase, the shift of the absorption edge is dominated by the evolution in ferroelastic shear strain.  

\begin{figure}[h!]
      \centering
      \includegraphics[width=\linewidth]{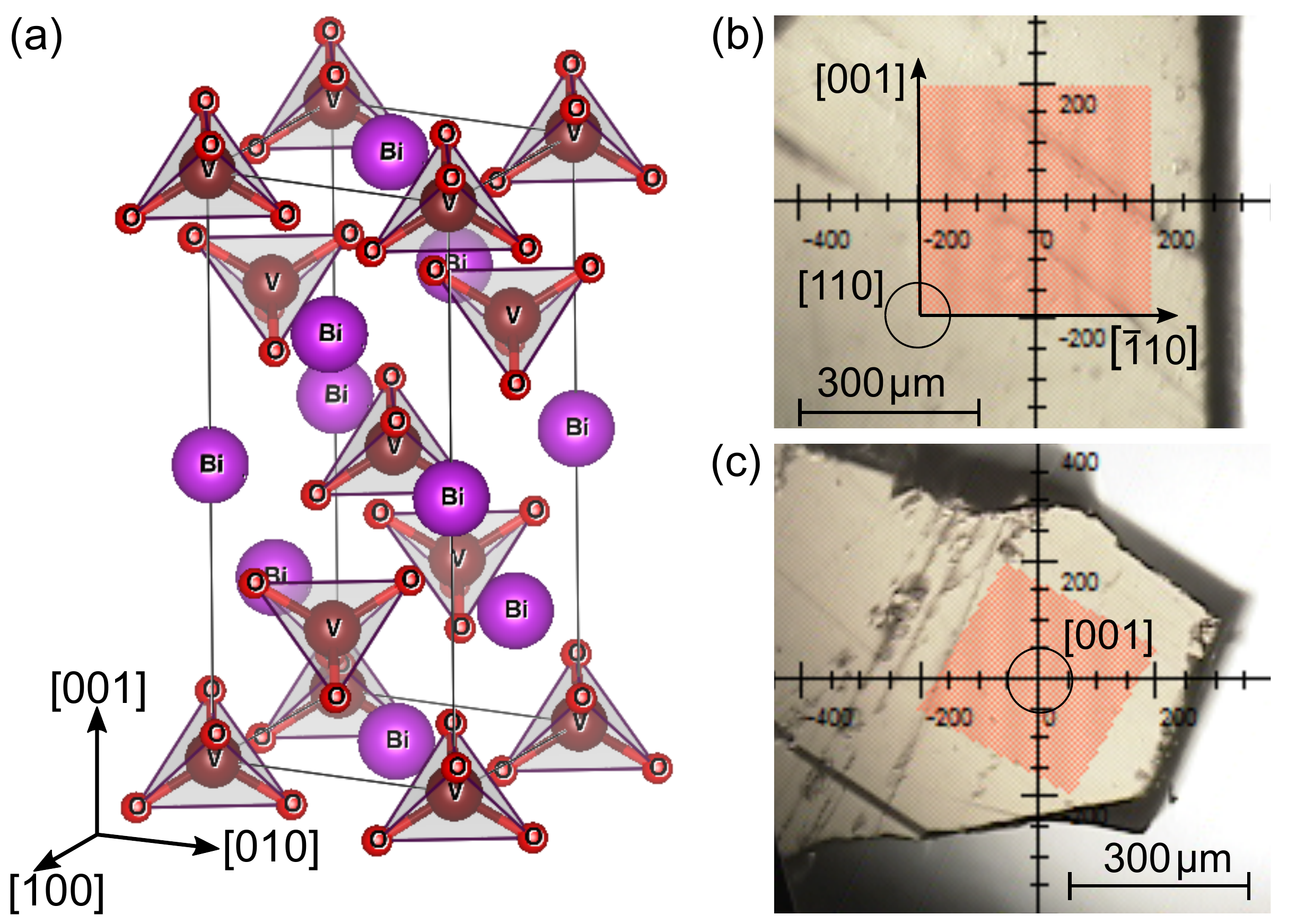}
  \caption{(a) Schematic representation of the tetragonal unit cell of BiVO$_4$. Optical microscopy images of (b) the $(110)$-oriented sample and (c) the $(001)$-oriented sample. For both samples, optical measurements were done with a vertical and a horizontal light polarization. The measurement areas are highlighted in red.}
  \label{fig:sample}
\end{figure}

Two samples were cut and polished from a larger single crystal pulled by the Czochralski-method. Fig.~\ref{fig:sample}(a) shows the tetragonal unit cell of scheelite BiVO$_4$; all forthcoming indications of orientations are given with respect to this tetragonal unit cell. Note that the tetragonal 4-fold axis becomes the monoclinic 2-fold axis, and thus remains the principal axis in both phases. The first sample has a surface orientation $(110)$ and a thickness of \SI{360}{\um}; the second sample has a $(001)$ oriented surface and a thickness of \SI{100}{\um}, see Figs.~\ref{fig:sample}(b) and \ref{fig:sample}(c). The transmittance and reflectance measurements were performed with a Jasco MSV-370 microspectrophotometer. The spectrophotometer operates a deuterium lamp for the UV and a halogen lamp for the visible and near infrared light range and collects the light using non-dispersive Schwarzschild objectives. The samples were placed in a Linkam THMS600 heating stage. Using an aperture, the measurement areas were set to $400\times 400$\,\si{\um\squared} and $300\times 300$\,\si{\um\squared} for the $(110)$-oriented sample and the $(001)$-oriented sample, respectively. In all cases, great care was taken to position the measurement area on the sample in homogeneous regions, i.e., free of domain walls or surface cracks. For all reflectance measurements, an aluminium mirror was used for calibration.

    \begin{figure}[h!]
      \centering
      \includegraphics[width=\linewidth]{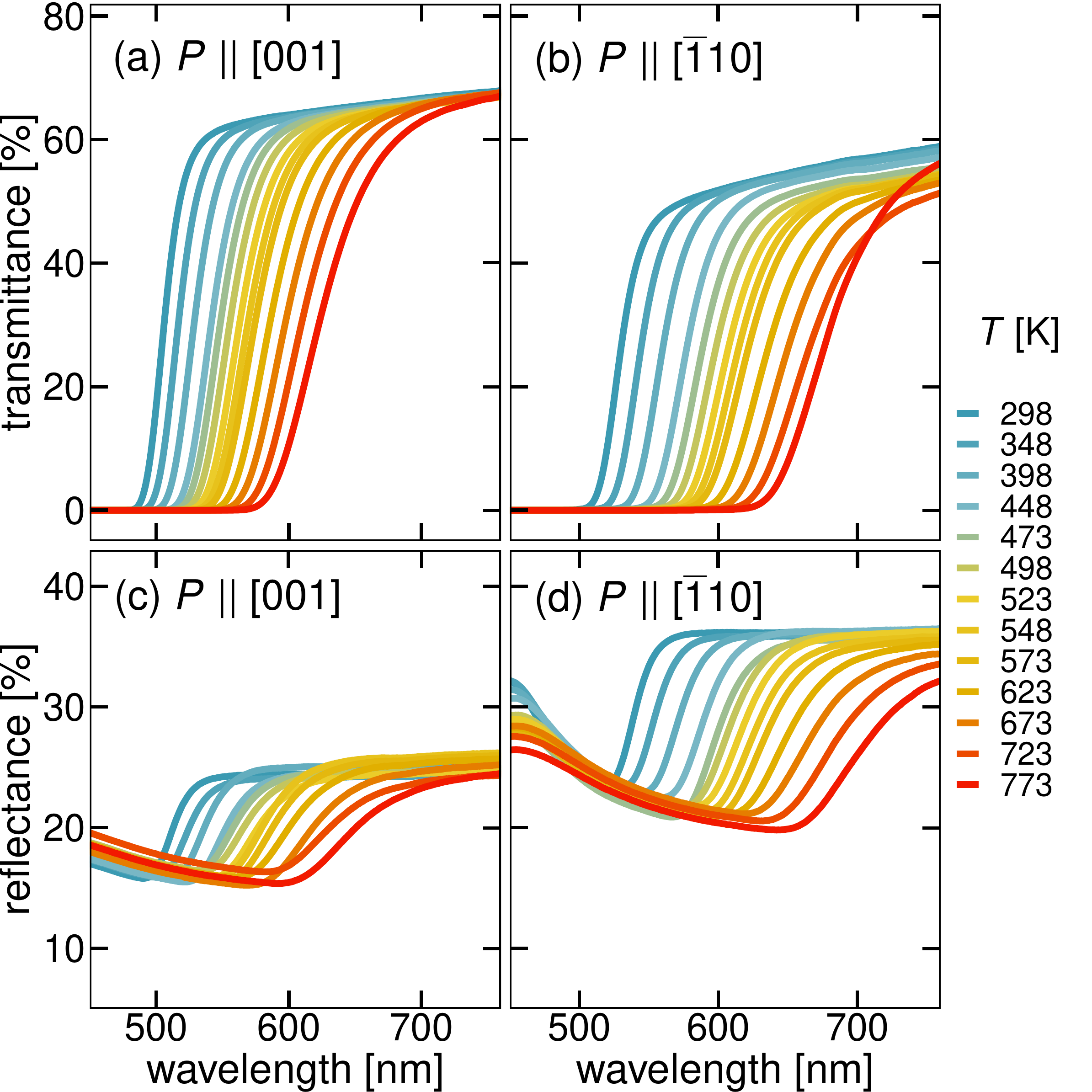}
    \caption{Transmittance spectra measured on the $(110)$-oriented sample (a) for $P \parallel [001]$ and (b) for $P \parallel [\bar{1}10]$ as a function of temperature ranging. (c) and (d) Reflectance spectra for $P \parallel [001]$ and $P \parallel [\bar{1}10]$ respectively.}
    \label{fig:dataRT}
    \end{figure}

\begin{figure}[h!]
  \centering
    \includegraphics[width=\linewidth]{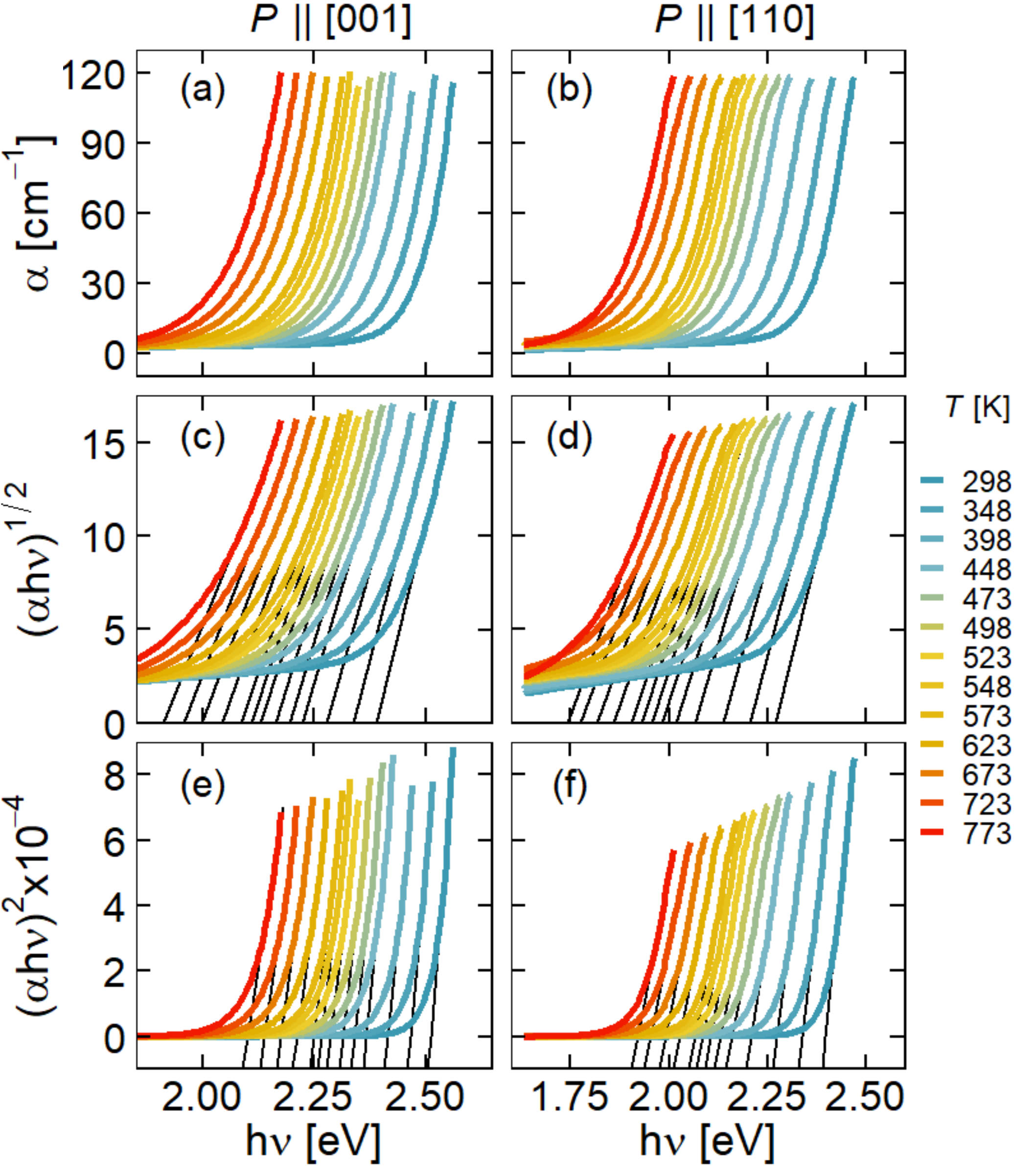}
    \caption{(a), (b) Optical absorption coefficient calculated from transmittance and reflectance data of the $(110)$-oriented sample for $P \parallel [001]$ and $P\parallel [\bar{1}10]$, respectively. (c), (d) and (e), (f) Tauc plots calculated for both polarization directions, assuming an indirect and a direct band gap, respectively.}
    \label{fig:alphaTauc}
\end{figure}

We first discuss the results for the $(110)$-oriented sample. Experiments were performed with two different directions of the linear light polarization $P$: first, with $P$ parallel to the crystallographic axis $[001]$ and second with $P \parallel [\bar{1}10]$, see Fig.~\ref{fig:sample}(b). 
The data are shown in Fig.~\ref{fig:dataRT}. For both light polarization directions, reflectance and transmittance data show a pronounced temperature dependence, with an absorption edge shifting to longer wavelengths as temperature increases. To investigate the nature of the band gap and its possible change with temperature, the absorption coefficient was calculated. In the spectral range of the fundamental absorption edge, the absorption coefficient $\alpha$ is approximated by 
\[
\alpha = \frac{1}{t}\ln\Big(\frac{1-R}{T}\Big),
\]
where $T$ is the transmittance and $R$ the reflectance and $t$ the sample thickness.\cite{demichelis_new_1987} Figs.~\ref{fig:alphaTauc}(a) and \ref{fig:alphaTauc}(b) show the absorption coefficient versus temperature for both polarization directions. 
The theory of interband optical absorption reveals that the absorption coefficient $\alpha$ varies with the photon energy according to the well-known relation $(\alpha h\nu)^n=A(h\nu-E_\text{opt})$, where \textit{A} is a constant and $E$\textsubscript{opt} is the optical absorption edge.\cite{pankove_optical_1975} Here, $n$ characterizes the nature of the transition process which dominates the optical absorption. We have $n=2$ for a direct band-to-band transition and $n=0.5$ for an indirect band-to-band transition. Usually, the value of $E$\textsubscript{opt} is determined by plotting $(\alpha h\nu)^n$ versus the photon energy $h\nu$ in a so-called Tauc plot. Hence, data are described by a straight line for $n=2$ in case of a direct band gap and for $n=0.5$ in case of indirect band gap. This allows to determine the nature of the band gap. $E$\textsubscript{opt} is given by the intercept of the linear fit with the photon-energy axis. Figs.~\ref{fig:alphaTauc}(c) and \ref{fig:alphaTauc}(d) show the Tauc plots for both light polarizations assuming indirect ($n=0.5$) and Figs.~\ref{fig:alphaTauc}(e) and \ref{fig:alphaTauc}(f) for direct ($n=2$) transition. In all four cases, the data can be described by a straight line making it difficult to conclude on the nature of the transition with this measurement alone. Considering the similar curve shape of the absorption coefficient at each temperature, there are no indications that first, the nature of the transition differs from one to the other light polarization and secondly, that the nature of the transition changes with temperature. Following the recent conclusions by Cooper et al. on the indirect nature of the band gap, we extract an energy for the absorption edge of \SI{2.27}{\electronvolt} for $P\parallel [\bar{1}10]$ and of \SI{2.39}{\eV} for $P \parallel [001]$ at room temperature.  
%

\begin{figure}[h!]
  \centering
  \includegraphics[width=1\linewidth]{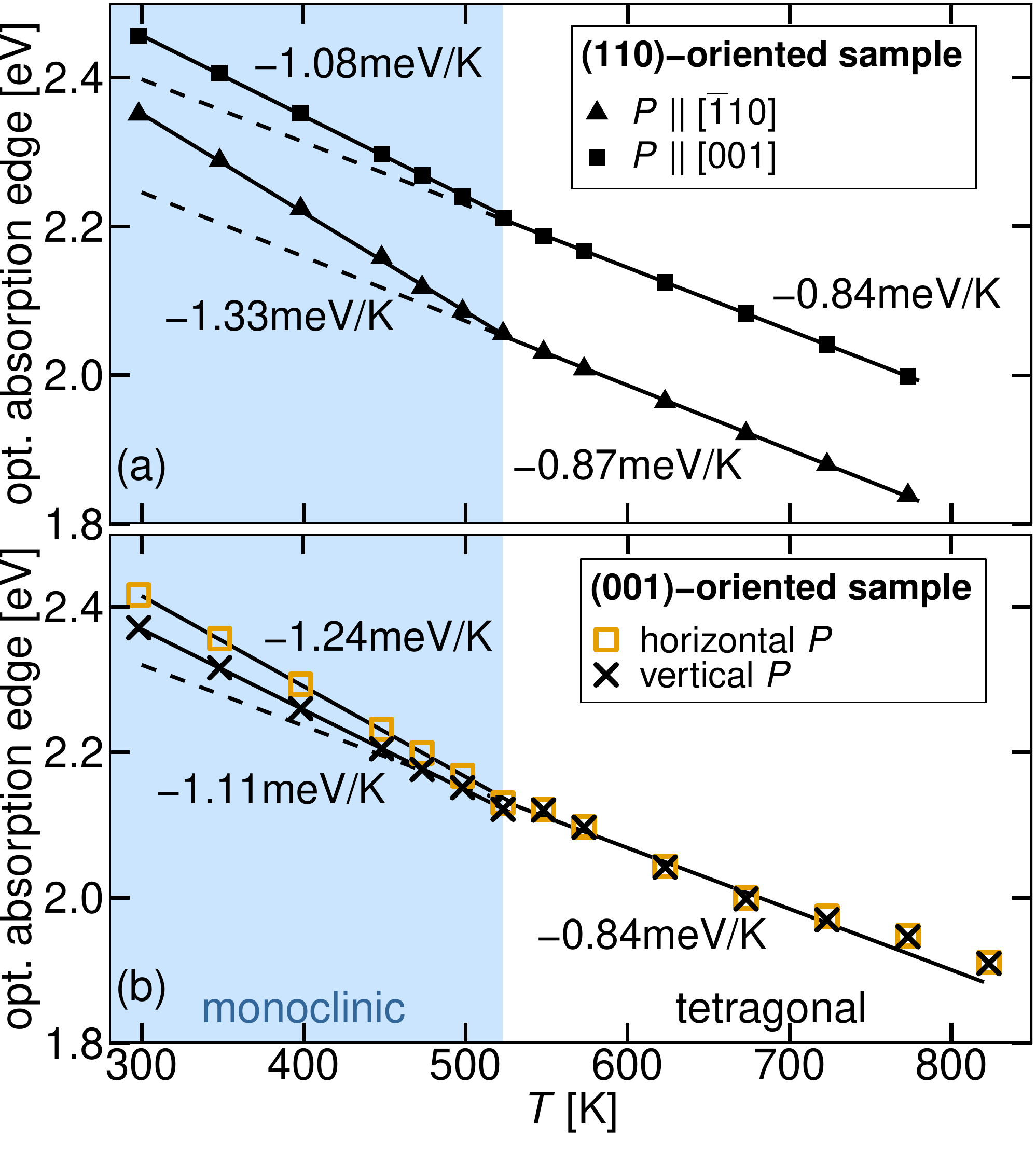}
  \caption{Temperature dependence of the optical absorption edge for (a) the $(110)$-oriented sample and (b) the $(001)$-oriented sample. The solid lines are fits to the data. The dashed lines are extrapolations of the high temperature trend into the low-temperature region.}
  \label{fig:absedge}
  \end{figure}

In a next step, we investigated the anisotropy of the optical absorption. To analyze the data without having to rely on any particular hypothesis on the nature of the band gap, we define the absorption edge as the inflexion point of the transmittance curve. We determined this point by fitting the first derivative of the transmittance curve with an asymmetric Gaussian function. This function was chosen because it produces a satisfactory fit for all data and provides a consistent analysis, although we do not claim any particular physical meaning for it. The results are shown in Fig.~\ref{fig:absedge}. In a first approximation, the evolution of the absorption edges was fitted linearly, with separate fits above and below $T_c$. Fig.~\ref{fig:absedge}(a) shows the absorption edge obtained for the $(110)$-oriented sample for both polarizations. We notice, first, that the absorption is clearly anisotropic in the tetragonal phase, with an energy difference of $\approx$\,\SI{0.2}{\eV} between the two polarizations. This difference remains constant for temperatures above $T$\textsubscript{c}, and the absorption edge decrease at the same rate of $\approx$\,\SI{-0.85}{\meV\per\K} for both light polarizations. Second, we observe a clear kink at $T$\textsubscript{c} in the temperature evolution of the absorption edge for both light polarizations. Below $T$\textsubscript{c}, in the monoclinic phase, the evolutions for both light polarizations remain linear, albeit with different slopes. This demonstrates that the absorption edge is sensitive to ferroelastic strain. These findings are robust even if alternative methods to extract the absorption are used. The corresponding plots are provided in the supplementary information. 

In the monoclinic phase, the optical absorption is expected to be anisotropic also in the plane perpendicular to the principal axis. To estimate the magnitude of this anisotropy, we performed a similar experiment with the $(001)$-oriented sample. We first measured the transmittance at room temperature, at a fixed wavelength of \SI{515}{\nm}, as a function of polarization direction (see supplementary information). We then positioned the crystal in such a way that the extrema of this curve were found for vertical and horizontal light polarizations (Fig.~\ref{fig:sample}(c)), and proceeded to the measurements over the full wavelength range. The difference in absorption edges measured in this configuration amounts to \SI{45}{\meV} at room temperature. Transmittance spectra were also collected with increasing temperature, and the corresponding absorption edges are shown in Fig.~\ref{fig:absedge}(b). The energy difference decreases with increasing temperature and vanishes at $T_\mathrm c$, as expected from the tetragonal symmetry. This should be considered as a first estimation, notably because we neglect here the effect of birefringence. A detailed study of the anisotropy in the monoclinic plane would require further experimental work. For the purpose of the present work, we simply observe that the anisotropy in this plane is comparatively small, so that the fundamental band gap is found for light polarized perpendicular to the principal axis, at room temperature and above. 

\begin{figure}[h!]
  \centering
  \includegraphics[width=1\linewidth]{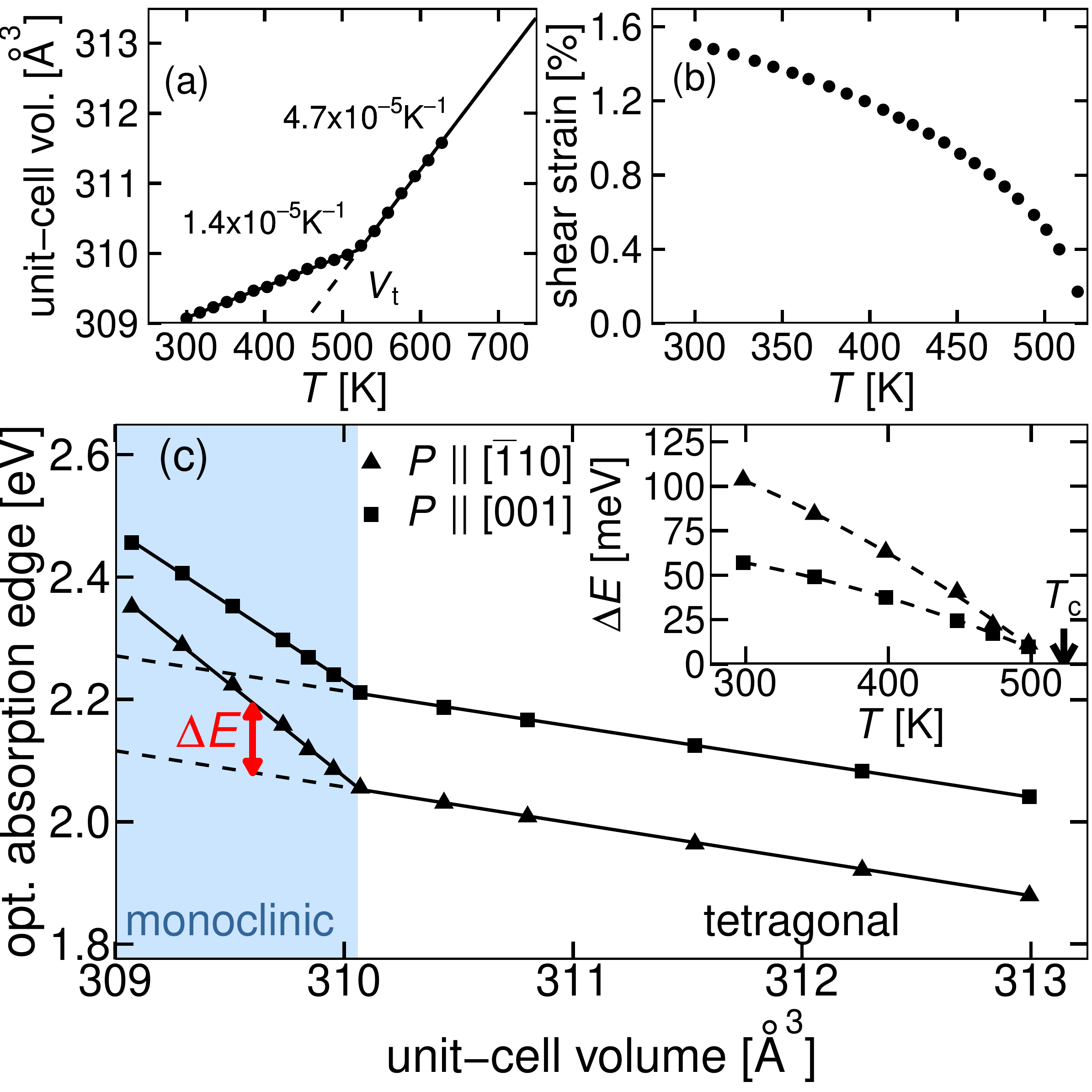}
  \caption{(a) Evolution of the unit-cell volume across the phase transition and (b) of the shear strain in the monoclinic phase, as discussed in the text. Data are taken from Ref.~\citenum{david_ferroelastic_1983_V}. (c) Position of the absorption edge for $P \parallel [001]$ and $P\parallel [\bar 110]$ as a function of the unit-cell volume. Inset: Difference in absorption edges $\Delta E$ associated to the phase transition as a function of temperature.}
  \label{fig:strain}
  \end{figure}

Having established that the optical absorption is clearly sensitive to the ferroelastic transition, it is insightful to decompose the observed variations of the absorption edges as a function of the different components of strain. For that purpose, we use the detailed knowledge of the volume, lattice constants and strains across the phase transition in Ref.~\citenum{david_ferroelastic_1983_V}, and focus here on the $(110)$-oriented sample. Fig.~\ref{fig:strain}(a) shows the evolution of the unit-cell volume across the phase transition and the corresponding thermal expansion coefficients. In the high-symmetry tetragonal phase, only thermal expansion is present and both absorption edges follow linear and parallel evolutions at an average rate of $\approx$\,\SI{-0.85}{\meV\per\K} for a volumetric thermal expansion coefficient of $\approx$\,\SI{4.7e-5}{\per\K} (details of the calculations are provided in the supplementary information). At the transition, the volume exhibits a kink and the thermal expansion is reduced to \SI{1.4e-5}{\per\K} below $T_c$. Assuming the same scaling, this would account for a change in absorption edge of $\approx$\,\SI{-0.25}{\meV\per\K}, which is only a small fraction of the total change shown in Fig.~\ref{fig:absedge}(a). We conclude that the ferroelastic shear strain dominates the evolution of the optical band gap in the monoclinic phase. This spontaneous shear strain develops in the $(ab)$ plane, has $B_g$ symmetry and is the primary order parameter of the transition.\cite{david_ferroelastic_1983_V} In Fig.~\ref{fig:strain}(b) we show  the temperature evolution of the total shear strain calculated from the $a$, $b$ and $\gamma$ lattice parameters as defined by Aizu~\cite{Aizu1970} and reported in Ref.~\citenum{david_ferroelastic_1983_V}. The respective influences of volume vs. shear strain to the optical absorption are best illustrated in Fig.~\ref{fig:strain}(c), where we show the absorption edges versus the unit-cell volume. The contribution of the simple volume change in the monoclinic phase is shown by the extrapolated dashed line and is clearly marginal. In the inset of Fig.~\ref{fig:strain}, we also show the difference in optical band gap $\Delta E$ caused by the phase transition as a function of temperature for both polarizations. In both cases, the evolution deviates from linearity and shows the beginning of a saturation behavior, which again is an indication that they scale predominantly with the shear strain.\cite{david_ferroelastic_1983_V} 

Finally, we discuss our observations in the light of the current theoretical understanding of the optical absorption in scheelite BiVO$_4$.\cite{das_investigation_2017, wadnerkar_density_2013,ding_density_2013,zhao_electronic_2011} Our experimental observations, as described above with our tentative Tauc plots, are consistent with the picture given in the introduction, at all temperatures up to \SI{773}{\K}. We confirm the strong anisotropy of the optical absorption. At all temperatures, and irrespective of the anisotropy in the $(001)$-plane, the lowest absorption edge is found for light polarized perpendicular to the principal axis. For a quantitative comparison of the band gap value(s) and the anisotropy, one should bear in mind the strong temperature dependence and therefore extrapolate experimental values down to \SI{0}{\K}. This extrapolation is not trivial, because the behaviour at low temperatures is non linear for various reasons: the strain dependence described above, and a possible saturation of the order parameter at low temperatures~\cite{Salje1991b}, but also the usual non-linear lattice anharmonicity and electron-phonon interactions.\cite{Vainshtein1999} The first effect can be taken into account by fitting a power law $\Delta E_\text{opt} = A (T_c - T)^\beta$ to the data below $T_c$ -- this yields an extrapolated value of \SI{2.62}{\eV} at \SI{0}{\K}. Saturation of the order parameter in BiVO$_4$ can be neglected, as demonstrated by birefringence studies.\cite{Wood1984} On the other hand, the other non-linear effects cannot be neglected a priori. In the supplementary information, we use physical arguments to show that they could lead to an additional lowering of the extrapolated value by as much as $\approx$\,\SI{0.3}{\eV}. In any case, the experimental value agrees well with the calculated value of \SI{2.47}{\eV} from Ref.~\citenum{das_investigation_2017}. On the other hand, we find a significant difference in the magnitude of the anisotropy between light polarized parallel and perpendicular to the principal axis: because of the influence of the shear strain, the difference between absorption edges is reduced to \SI{90}{\meV} at \SI{0}{\K}, which is much weaker than the theoretical value\cite{das_investigation_2017} of \SI{0.5}{\eV}. In light of the sensitivity to shear strain described above, we hypothesize that these disagreements find their origin in the accuracy with which the calculation captures the small monoclinic distortion of the scheelite structure. A better quantitative predictions will require a particular attention to these structural aspects. 

In summary, the optical absorption in bismuth vanadate has been measured over a wide temperature range, and the importance of its ferroelastic transition has clearly been established. The band gap of \SI{2.27}{\eV} extracted at room temperature is in good agreement with the band gap values published previously. While our data do not directly provide conclusive support for the indirect nature of the band gap, they do show that its nature remains unchanged in the investigated temperature range. We find that the high-symmetry tetragonal phase exhibits a constant anisotropy in optical absorption of \SI{0.2}{\eV}, depending whether the light is polarized parallel or perpendicular to the principal axis. The fundamental band gap was found for light perpendicular to the principal axis, consistent with expectations from DFT calculations, pointing to the possibility to  activate different band-to-band transitions by playing with the polarization of the light. We show that the position of the absorption edge of bismuth vanadate is very sensitive to the onset of the spontaneous shear strain which develops in the low-symmetry phase, dominates the evolution and enhances the thermochromic behaviour. These results suggest that the optimization of the optical properties of BiVO$_4$-based light absorbers would benefit from a careful control of crystallite orientation. The sensitivity to strain could also be exploited via interface effects, epitaxy, nano-size effects or cation substitution.

\section*{Supplementary Material}
See supplementary material for a description of alternative methods to extract the absorption edge from optical data, details on the positioning of the (001)-oriented sample, details on the calculation of the thermal expansion and a discussion on the non-linear evolution of the absorption edge at low temperatures.

\begin{acknowledgments}
CH acknowledges funding from the Fond National de la Recherche under Project PRIDE/15/10935404. MF and JL acknowledge support by the ETH Grant no. ETH-28 14-1. MF and MW are grateful for financial support from the SNSF (Grant No. 200021\_178825). The authors also gratefully acknowledge M.~Glazer (University of Oxford) for discussions and valuable insights into early studies of BiVO$_4$ single crystals. \\ 
\end{acknowledgments}

The data that support the findings of this study are available from the corresponding author upon reasonable request.
\medskip


\end{document}